\documentclass[aps,prb,twocolumn,showpacs,preprintnumbers,amsmath,amssymb,superscriptaddress]{revtex4}

\usepackage{graphicx}
\usepackage{dcolumn}

\newcommand{\be}{\begin{equation}}
\newcommand{\ee}{\end{equation}}
\newcommand{\bea}{\begin{eqnarray}}
\newcommand{\eea}{\end{eqnarray}}
\newcommand{\bm}[1]{\mathbf{#1}}

\newcommand{\lp}{\left(}
\newcommand{\rp}{\right)}

\def \cP{{\cal P}}
\def \cQ{{\cal Q}}
\def \cH{{\cal H}}

\def \cD{{\cal D}}

\newcommand{\ty}[1]{\mbox{\tiny #1}}

\begin{document}

\title{Hydrodynamic theory of transport in doped graphene}

\author{R. Bistritzer and A.H. MacDonald}
\affiliation{Department of Physics, The University of Texas at Austin, Austin Texas 78712\\}

\date{\today}

\begin{abstract}

We study non-linear {\em dc} transport in graphene using a hydrodynamic approach
and conclude that in clean samples the drift velocity saturates at a weakly density-dependent value
$v_{sat} \sim 10^7$ cm/s. We show that saturation results from the interactions between graphene's Dirac quasi-particles and both acoustic and optical phonons.
Saturation is accompanied by substantial electron heating and is not reached at
realistic driving fields in moderately or strongly disordered samples.
We find that it is essential to account for interactions among graphene's Dirac quasi-particles,
which increase the linear response resistivity at high temperatures or low densities.

\end{abstract}


\maketitle

\noindent
\section{Introduction}

The unique electronic and thermal properties of graphene two-dimensional electron systems
make them promising as potential building blocks  for future electronic devices.\cite{reviewMacDonald,reviewNeto}
The gapless Dirac-like spectrum of graphene presents an obstacle to logic device applications,
but is also partly responsible for large quasiparticle velocities which are advantageous in
analog and radio-frequency devices.\cite{meric}
The feasibility of these applications
is dependent mainly on graphene's non-linear electrical response properties which we address
in this paper using a hydrodynamic approach.

The response of graphene to an external electric field is determined by the interaction of its Dirac quasi-particles with
impurities, with phonons, and with each other.\cite{tan,Fratini,FuhrerDisorder,Morozov,stauberBoltzmann,Hwang,stauberFiniteV,Adam}
For samples on substrates, the resistivity contribution  due to scattering from the phonon modes of graphene is unimportant
in the linear response regime even at room temperature.  The linear resistivity is generally believed to be limited
by elastic scattering off Coulomb impurities.\cite{ando,Nomura,tan}
Even in annealed suspended graphene sheets, which have dramatically weakened elastic scattering, the phonon-limited resistance is
irrelevant in currently available samples because it is small
compared to the quantum resistance.\cite{Bolotin,Du}  At high-fields, however,
electron-phonon (e-ph) interactions are essential for two reasons.
First, the phonon modes are the only dissipative channels through which the electrons can lose the energy that they acquire
by flowing in the presence of a high electric field.
Second, the electronic temperature rises and the drift velocity $u$
increases with the electric field; both changes enhance e-ph scattering, making
it much more efficient for momentum relaxation.
Extrinsic phonons localized near the substrate surface can play an important role even in the linear regime.\cite{Fratini,FuhrerSubstarteSiO2}
However the significance of these phonons is sensitive
to the specific experimental system, in particular to the type of substrate and its distance from the graphene sheet.
We therefore do not account explicitly for these extrinsic phonon modes
and instead limit ourselves to an explanation of how they may be straightforwardly incorporated
when the parameters appropriate for a particular experimental system are known.

One very important consequence of the high electronic temperature in
the non linear regime is enhanced electron-electron (e-e) scattering.
As recent spectroscopy measurements\cite{deHeer,Spencer} demonstrate, e-e scattering is the dominant scattering mechanism at high electronic temperatures.
High temperatures are inevitable in the non linear regime because of Joule heating. In the linear regime
e-e interactions dominate either when the sample is hot or when it is disorder-free.
When it is dominant, the influence of e-e scattering on transport can not be considered perturbatively.
In this paper we use a theoretical approach to non-linear {\em dc} transport
which exploits rapid e-e collisions
by using a hydrodynamic theory.  One important advantage of this theory is its simplicity and physical transparency.
As we explain below, in a hydrodynamic theory the non-equilibrium system is characterized by only three
parameters: the chemical potential $\mu$, the electronic temperature $T_e$ and the drift velocity $\bm{u}$.

Our paper is organized as follows. In section \ref{section_hydrodynamic} we derive the hydrodynamic equations.
Using these equations we study the linear {\em dc} transport in graphene in section \ref{section_linear_response}.
We then consider the non-linear case in section \ref{sec_non_linear}.
We first focus on a clean system and then discuss the role of disorder. Finally we summarize our findings in section \ref{sec_summary}.
As we explain below, the hydrodynamic description breaks down in the neutral regime when $|\mu|/T_e \ll 1$. We therefore
restrict our study to doped systems.
Because the Dirac model for graphene is perfectly particle-hole symmetric we can
restrict our attention to electron-doped systems without loss of generality.

\section{Hydrodynamic theory for doped graphene       \label{section_hydrodynamic}}

The Boltzmann theory provides a simple but faithful description of transport in many electronic systems.
In graphene, the validity of this semiclassical transport theory is well established for the doped systems we study.
The backbone of Boltzmann transport theory is the distribution function $f_{\bm{k}\alpha}(\bm{r},t)$ defined
as the occupation probability of the Bloch state in band $\alpha$ with crystal momenta $\bm{k}$ at position $\bm{r}$ and time $t$.
All physical quantities can be expressed in terms of $f$. The distribution
function is determined by requiring that it satisfies the Boltzmann equation:
\be
\lp \partial_t + \bm{v_{\bm{k}\alpha}} \cdot \nabla +  e \bm{E} \cdot \nabla_{\bm{k}} \rp  f_{\bm{k}\alpha}(\bm{r},t) = S_{eL} + S_{ee}       \label{boltzmann_eq}
\ee
where $\bm{v}_{\bm{k}\alpha}$ is the band velocity, $\bm{E}$ is the electric field,
$S_{eL}$ is the electron-lattice collision integral which accounts for electron scattering by phonons and disorder, and $S_{ee}$ is the e-e
collision integral which accounts for e-e scattering.

Further simplification of the theory is possible in the hydrodynamic regime when the e-e scattering time $\tau_{ee}$ is considerably
shorter than the e-ph scattering time $\tau_{ph}$ and the impurity scattering time $\tau_i$, \emph{i.e.} when
\be
\tau_{ee} \ll \tau_{ph} , \tau_{i}.         \label{hydrodynamic_condition}
\ee
This separation of time scales implies that, to leading order in both $\tau_{ee}/\tau_{i}$ and $\tau_{ee}/\tau_{ph}$,
rapid e-e collisions are able to establish a drifting Fermi distribution function\cite{Levinson}
\be
f^{\ty H}_{\bm{k}\alpha}(\bm{r},t) = \left[ \exp\lp\frac{\epsilon_{\bm{k}\alpha} - \bm{u(\bm{r},t) \cdot k} - \mu(\bm{r},t)}{T_e(\bm{r},t)}\rp + 1 \right]^{-1},    \label{fk}
\ee
where $\epsilon_{\bm{k}\alpha}$ is the energy dispersion of band $\alpha$.
This form of distribution function satisfies $S_{ee}(f) = 0$.  The drift velocity $\bm{u}$ can be non-zero because
e-e scattering does not relax momentum.  The existence of the three hydrodynamic functions $T_e,\bm{u}$ and $\mu$ is a direct consequence of the conservation of
energy, momentum, and particle-number in e-e collisions.

Specializing to {\em dc} transport in graphene, the Boltzmann equation reduces to
\be
e \bm{E} \cdot \nabla_{\bm{k}} f^{\ty H}_{\bm{k}\alpha} = S_{eL} + S_{ee}.      \label{boltzmann_eq_DC}
\ee
In Eq.(\ref{boltzmann_eq_DC}) the hydrodynamic functions are independent of time and position, and
$\epsilon_{\bm{k}\alpha} = \alpha v k$ with $v$ being the band
velocity of graphene.
In what follows we interchangeably use $\alpha=c,v$ and $\alpha=+,-$ to label the conduction and valence bands of graphene.

The three hydrodynamic parameters give a full description of the non-equilibrium state. Their values
are fixed by the aforementioned conservation laws.
Multiplying Eq.(\ref{boltzmann_eq_DC}) by $\bm{k}$, summing over it and the band index, and using $\sum_{\bm{k}\alpha}\bm{k}S_{ee}=0$ implied by the conservation of momentum
in an e-e scattering event we obtain the force balance equation:
\be
e n \bm{E} = \cP               \label{DC_momentum}
\ee
where
\be
\cP = -g \sum_{\bm{k}\alpha} \bm{k} S_{eL}\lp f^{\ty H}_{\bm{k}\alpha} \rp.        \label{P}
\ee
Here  $g=4$ accounts for the spin and valley degeneracies in graphene. The equality in Eq.(\ref{DC_momentum}) expresses the steady state balance
between the momentum acquired by the charge carriers due to the electric field and the momentum lost by scattering off phonons and impurities.

Similarly by multiplying Eq.(\ref{boltzmann_eq_DC}) by $\epsilon_{\bm{k}\alpha}$, summing over momenta and over the band index and using
$\sum_{\bm{k}\alpha} \epsilon_{\bm{k}\alpha} S_{ee} = 0$ we obtain the energy balance equation:
\be
e n \bm{E \cdot u} = \cQ      \label{DC_energy}
\ee
where
\be
\cQ = -g \sum_{\bm{k}\alpha} \epsilon_{\bm{k}\alpha} S_{eL}\lp f^{\ty H}_{\bm{k}\alpha} \rp.        \label{Q}
\ee
$\cQ$ is positive when $T_{e} > T_{L}$.  In {\em dc} transport the energy gained by carriers due to drift in an electric field must be balanced by energy
lost to the phonon bath.

Finally, the third equation necessary to fix the values of the hydrodynamic variables
\be
n = g \sum_{\bm{k}} \left[ f^{\ty H}_{\bm{k}c} - \lp 1 - f^{\ty H}_{\bm{k}v} \rp \right]    \label{number_eq}
\ee
follows from number conservation.
Eqs.(\ref{DC_momentum},\ref{DC_energy}) and (\ref{number_eq}) determine the
hydrodynamic parameters $\mu$, $T_e$ and $\bm{u}$ , given the values for the electric field $\bm{E}$,
the lattice temperature $T_{\ty L}$ and the density $n$.

In subsequent sections we use the hydrodynamic equations to study {\em dc} transport in graphene.
We start in the next section by considering the linear electrical response of graphene,
before turning in the following section to the full non-linear response.

\section{Linear response \label{section_linear_response}}

When high currents are driven through a graphene sheet it is heated \emph{i.e.} $T_e > T_{\ty L}$.
The chemical potential is then reduced relative to its
equilibrium value to maintain a fixed electronic density.
Symmetry considerations imply that inversion of the electric field inverts the drift velocity, but
renders $T_e$ and $\mu$ unchanged. Therefore
in the linear response regime, {\em i.e.} to first order in $E$, $T_e$ and $\mu$ retain their equilibrium values and the energy balance equation (\ref{DC_energy})
and the number equation (\ref{number_eq}) are satisfied identically. The drift velocity $\bm{u}$ follows from the momentum balance
equation (\ref{DC_momentum}).

At physically relevant temperatures, momentum loss in graphene is thought to be primarily due to long range Coulomb scatterers and secondarily due to interactions of electrons
with longitudinal acoustic phonons.  We find, in accord with experiment, that the momentum loss rate due to the energetic intrinsic optical phonons is negligible.
Above the  Bloch--Gr$\ddot{u}$neisen temperature $T_{\ty{BG}} \approx 2 c \sqrt{\pi n}$ scattering by acoustic phonons is quasi-elastic
due to the large mismatch between the sound velocity $c$ and graphene's band velocity $v$.
For elastic scattering
\be
\cP = g\sum_{\bm{k,p}\alpha} \bm{k} \lp f^{\ty H}_{\bm{k}\alpha} - f^{\ty H}_{\bm{p}\alpha} \rp W_{\bm{k,p}}     \label{P_linear}
\ee
where $W_{\bm{k,p}}$ is the transition rate between states $\bm{k}$ and $\bm{p}$.
Expanding $f^{\ty H}$ to linear order in $\bm{u}$ and using
\be
\sum_\bm{p} \cos\theta_\bm{p} \ W_{\bm{k,p}} = \cos\theta_\bm{k} \sum_\bm{p} \cos\theta \ W_{\bm{k,p}},
\ee
where $\theta=\theta_{\bm{k}} - \theta_{\bm{p}}$ is the relative angle between the incoming and outgoing momenta, we find that
\be
e n \bm{E} = -g \frac{\bm{u}}{2} \int \frac{k^3}{2\pi} \frac{1}{\tau_k v_k} \partial_k \lp f_{kc}^{(0)} - f_{kv}^{(0)} \rp   \label{u}
\ee
from which the drift velocity $\bm{u}$ readily follows. Here $f_{k\alpha}^{(0)}$ is the equilibrium Fermi distribution function, $v_k = \partial_k \epsilon_k$ is the band velocity
and $ \tau_k^{-1} = \sum_{\bm{p}} (1-\cos\theta) W_{\bm{k,p}} $ is the elastic (transport) scattering rate. The resistivity
\be
\rho_{ee} = -\frac{g}{2e^2 n^2} \int \frac{k^3}{2\pi} \frac{1}{\tau_k v_k} \partial_k \lp f_{kc}^{(0)} - f_{kv}^{(0)} \rp.      \label{rho_ee}
\ee
follows from Eq.(\ref{u}) and from the expression for the current
\be
\bm{I} = e \sum_{\bm{k}\alpha} \bm{v_{k\alpha}} f^{\ty H}_{\bm{k}\alpha} = e \bm{u} n.     \label{I}
\ee

To illustrate the influence of e-e interactions on the resistivity we compare $\rho_{ee}$ to
\be
\rho_{\ty 0} = -\frac{2}{ge^2} \left[ \int \frac{k dk}{2\pi} v_k \tau_k \partial_k \lp f_{\bm{k}c}^{(0)} + f_{kv}^{(0)} \rp \right]^{-1}            \label{rho_0}
\ee
the resistivity obtained directly from the Boltzmann equation when e-e interactions are neglected.
At zero temperature the hole density in the valence band vanishes and the
resistivity is not modified by e-e interactions.
At finite temperatures, however, $\rho_{ee}$ is always larger than $\rho_{\ty 0}$.
Note that the resistivity expressions, (\ref{rho_ee}) and (\ref{rho_0}), assume only quasi-elastic scattering and
isotropy and are valid irrespective of the
energy dispersion.

\begin{figure}[h]
\includegraphics[width=0.95\linewidth]{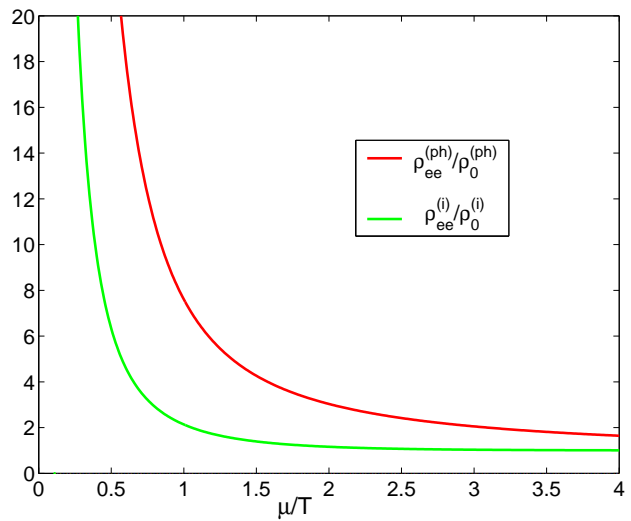}
\caption{Increase in resistivity due to e-e interactions.}
\label{fig:H}
\end{figure}

It is instructive to express the resistivity in graphene  as the sum of the residual resistivity $\rho^{(i)}$ and the acoustic phonon induced resistivity $\rho^{(ph)}$.
The residual resistivity follows from Eqs.(\ref{rho_ee},\ref{rho_0}) by setting $\tau_k^{(i)} = v k/u_0^2$ for the momentum relaxation time associated with Coulomb scatterers.
Here $u_0^2 = n_i \lp \pi e^2/\varepsilon \rp^2 $ where $n_i$ is the impurity concentration
and $\varepsilon$ is the dielectric function.\cite{reviewNeto} It follows from Eqs.(\ref{rho_ee},\ref{rho_0}) that
\be
\rho_{ee}^{(i)} =\rho_{0}^{(i)} \lp \frac{n_e + n_h}{n} \rp^2.      \label{rho_ee_impurities}
\ee
where $\rho_{\ty 0}^{(i)} = u_0^2/\left[ e^2 v^2(n_e + n_h) \right]$.
The dependence of $\rho^{(i)}$ on the number of electrons $n_e$ and the number of holes $n_h$
is qualitatively changed due to e-e interactions. As the temperatures is raised both $n_e$ and  $n_h$ increase while the total density $n=n_e-n_h$ remains fixed.
Thus e-e interactions change $\rho^{(i)}$ from being a monotonic decreasing function of temperature to a monotonic increasing function of temperature.
The ratio $\rho_{ee}^{(i)}/\rho_{0}^{(i)}$ is plotted in figure \ref{fig:H}.

As the temperature is increased the influence of phonon scattering on the resistivity becomes more important.
The momentum relaxation time associated with acoustic phonons is $\tau_k^{(ph)} = v/k C(T_{\ty L})$ where
$ C(T) = \cD^2 T/2\rho c^2  $, and $\cD$ is the deformation potential \cite{stauberBoltzmann}. It therefore
follows from Eqs.(\ref{rho_ee}) and (\ref{rho_0}) that above the Bloch--Gr$\ddot{u}$neisen temperature
\be
\rho^{(ph)}_{ee} = \rho^{(ph)}_0 \cH\lp \mu/T \rp.     \label{rho_acoustic}
\ee
Here $\rho^{(ph)}_{\ty 0} = \pi C(T)/e^2 v^2$ is the phonon induced resistivity in the absence of e-e interactions and
\be
\cH(z) = \frac{\int_0^\infty x^3 dx  \left[ h(x,z) + h(x,-z) \right] }{\left\{ \int_0^\infty x dx \left[ h(x,z) - h(x,-z) \right] \right\}^2}
\label{H}
\ee
where $h(x,z) = [\exp(x-z)+1]^{-1}$. The function $\cH$ is plotted in figure \ref{fig:H}.
As expected from our previous discussion $\cH$ approaches unity in the  $\mu/T \to \infty$ limit. In the opposite limit
$\cH \approx 5.91 \lp T/\mu\rp^2$.

Why is it that strong e-e interactions increase the resistivity?  When e-e interactions are neglected electrons and holes
contribute additively to the current. For nearly neutral systems, this
Boltzmann theory property explains the increase of the conductivity as the temperature is raised\cite{Bolotin,Du}.
In the opposite hydrodynamic limit, strong e-e interactions enforce a common drift velocity for all momenta and for both valence and conduction bands.
The currents in the two bands then flow in opposite directions, resulting in a smaller net current or equivalently in a higher resistivity. The difference between the
magnitudes of the two counter-flowing currents decreases as $T/|\mu|$ is increased.  These counterflow currents are reminiscent of the
Coulomb drag effect in electron-hole bilayers\cite{Sivan} when the two layers are contacted simultaneously.
The hydrodynamic description breaks down for an electron doped system once the momenta acquired by
the valence band holes due to the electric field exceeds the e-e induced momenta transfer between the bands.
Therefore our theory is invalid in the neutral regime when $|\mu|/T \ll 1$.

We now turn to study the non linear {\em dc} electric response of graphene.

\section{Non linear response     \label{sec_non_linear}}

Strong electric fields drive the system out of equilibrium. Far from equilibrium
the chemical potential is reduced relative to its equilibrium value and
the electronic temperature is higher than the lattice temperature. To find the
hydrodynamic parameters which characterize the non-equilibrium state,
we solve the three coupled hydrodynamic equations (\ref{DC_momentum},\ref{DC_energy})
and (\ref{number_eq}). We start by simplifying the general expressions (\ref{P}) and (\ref{Q}) for the momenta loss $\cP$ and energy loss $\cQ$.
In the following we assume that $\bm{E} \parallel \bm{\hat{x}}$.

\subsection{Momentum loss}

The momentum loss rate of the electronic system
\be
\cP = g\sum_{\bm{k\alpha}} k_x \sum_{\bm{p}\gamma} \left[ f^{\ty H}_{\bm{k}\alpha}(1-f^{\ty H}_{\bm{p}\gamma}) W_{\bm{k} \bm{p}}^{\alpha\gamma} - (\bm{k}\alpha \leftrightarrow \bm{p}\gamma) \right]  \label{Pgeneral}
\ee
has contributions due to scattering by disorder and by acoustic and optical phonons.
To calculate $\cP$ we must evaluate the transition rate $W$ and the momentum loss rate
for each scattering mechanism.

For elastic collisions expression (\ref{Pgeneral}) can be simplified:
\bea
\cP^{el} &=& g\sum_{\bm{kp}\alpha} k_x( f^{\ty H}_{\bm{k}\alpha} - f^{\ty H}_{\bm{p}\alpha}) W_{\bm{kp}}^{\alpha\alpha}        \label{P_el} \\
&=& g \sum_{\bm{k}} \frac{k}{\tau_k} \cos\theta_k \left[ f^{\ty H}_{\bm{k}c} - (1 - f^{\ty H}_{\bm{k}v}) \right].      \nonumber
\eea
Obviously, elastic scattering can not induce inter-band scattering in graphene .
The momentum loss rate due to scattering off Coulomb impurities,
\be
\cP_i = \frac{g \beta u_0^2}{v(1-\beta^2)^{3/2}} \sum_{\bm{k}} \left[ f_{\bm{k}c}^{(0)} + \lp 1 - f_{\bm{k}v}^{(0)} \rp \right],      \label{P_i}
\ee
is obtained by substituting  $\tau_k^{(i)}$ in Eq.(\ref{P_el}) and integrating over the angle. Here $\beta=u/v$ and $f^{(0)}$ corresponds to the Fermi function ($\bm{u}=0$)
with the chemical potential given by its non-equilibrium value. Similarly by substituting $\tau_k^{(ph)}$ in (\ref{P_el}) we obtain
\be
\cP_a = \frac{g \beta C(T_{\ty L})(4+\beta^2)}{2v(1-\beta^2)^{7/2}} \sum_{\bm{k}} k^2 \left[ f_{\bm{k}c}^{(0)} + \lp 1 - f_{\bm{k}v}^{(0)} \rp \right],     \label{Pa}
\ee
the momentum loss rate contribution from acoustic phonon scattering.

Electronic collisions with optical phonons \cite{andoOptical,Calandra,Piscanec,Yan}are highly inelastic. To obtain $\cP_o$, the contribution of a single phonon branch to the momentum loss, we substitute
\be
W_{\bm{k}\bm{p}}^{\alpha\gamma}=\sum_q w_q^{\alpha\gamma}\left[ (N_q+1)\delta(\epsilon_{kp}^{\alpha\gamma}-\omega_q)+N_q\delta(\epsilon_{kp}^{\alpha\gamma}+\omega_q)\right]
\label{W}
\ee
in Eq.(\ref{Pgeneral}). Here $N_q = N(\omega_q)$ is the Bose distribution function
evaluated at the phonon energy $\omega_q$, $w_q^{\alpha\gamma}$ is the golden rule expression for the transition rate from band $\alpha$ to band $\gamma$ via an interaction with a phonon
of momenta $\bm{q}$ and $\epsilon_{kp}^{\alpha\gamma} = \epsilon_{k\alpha}-\epsilon_{p\gamma}$.
We consider the two optical phonon branches. Conservation of momentum restricts the phonon momenta to be either near the zone center $\Gamma$ point or near the zone
edge $K$ point. Near the $\Gamma$ point both the longitudinal and transverse optical phonons couple to the electrons whereas near
the $K$ point it is mainly the $A'_1$ transverse phonon mode that causes inter-valley transitions.
Since the typical phonon momentum measured from the relevant symmetry point is small compared to the zone boundary momenta we approximate the phonon
energy by a constant: $\omega_\Gamma = 196meV$ near the $\Gamma$ point and $\omega_K = 167meV$ near the $K$ point. Furthermore we approximate
$w_q^{\alpha\gamma}$ by a momenta independent constant $g_\Gamma^2\approx 2 v/(a^2\sqrt{2\rho \omega_{0\Gamma}})$ for the zone-center phonons and
$g_K^2 \approx 2 g_\Gamma^2$ for the zone boundary phonons \cite{Rana}.

\subsection{Energy loss}

Since collisions with impurities are elastic, only phonons contribute to the energy loss rate
\be
\cQ = g \sum_{\bm{k}\alpha\bm{p}\gamma} \epsilon_{kp}^{\alpha\gamma} f^{\ty H}_{k\alpha}(1-f^{\ty H}_{p\gamma}) W_{\bm{k} \bm{p}}^{\alpha\gamma}.   \label{Qgeneral}
\ee
The transition rate $ W_{\bm{k} \bm{p}}^{\alpha\gamma}$ from state $\bm{k}\alpha$ to state $\bm{p}\gamma$ is given by Eq.(\ref{W}).

We first consider the energy loss due to acoustic phonons. Electronic transitions are induced only by the longitudinal mode for which
$ w_q^{\alpha\gamma} = \pi \cD^2 q^2 (1+\alpha\gamma\cos\theta)/2\rho\omega_q $ \cite{Suzuura}.
Here $\theta=\theta_k-\theta_p$ is the angle between the incoming and outgoing momenta, $\cD$ is the deformation potential, $\rho$ is the mass
density of graphene and $\omega_q=c q$ where $c$ is the sound velocity. We evaluate $\cQ_a$ to leading order in $c/v \ll  1$. Since the transitions
are elastic to zeroth order in $c/v$, the leading term of $\cQ_a$ is of order of $(c/v)^2$.

It is instructive to write $\cQ_a$ as a sum of $\cQ_a^{ind}$, the contribution to $\cQ_a$ due
to induced transitions, and the spontaneous emission contribution $\cQ_a^{sp}$.
The calculations are described in appendix \ref{app_acoustical_phonons_Q}. We find that
\be
\cQ_a^{ind} = -g \frac{\cD^2 T_{\ty L}}{2\rho v^2} \frac{2+3\beta^2}{(1-\beta^2)^{7/2}} \sum_{\bm{k}} k^2 \left[ f_{\bm{k}c}^{(0)} + \lp 1 - f_{\bm{k}v}^{(0)} \rp \right]. \label{Qa_ind}
\ee
As the lattice temperature increases the acoustic mode population increases rapidly and the rate of energy loss due to induced transitions increases.
On the other hand,
\be
\cQ_a^{sp} = g\frac{\cD^2}{4\rho v} \sum_\alpha \int \frac{k^4 dk}{2\pi}\left[ I_0^\alpha(1-I_0^\alpha) + (I_2^\alpha)^2 \right]      \label{Qa_sp}
\ee
with
\be
I_n^\alpha = \int \frac{d\theta}{2\pi} \cos(n\theta) f_{\bm{k}\alpha}     \label{I_definition}
\ee
depends on $T_e$ but is independent of the lattice temperature.
At equilibrium the energy gain due to the induced transitions is exactly compensated by the energy loss due to the spontaneous emission.
However, when the system is out of equilibrium $\cQ_a^{sp} > \cQ_a^{ind}$ resulting in a net rate of energy loss by the electronic system.

The energy loss due to the interaction of electrons with optical phonons $\cQ_o$ is evaluated using Eqs.(\ref{W},\ref{Qgeneral}).
In expression (\ref{W}) for $W_{\bm{kp}}^{\alpha\gamma}$ we make the same approximations made above to evaluate $\cP_o$; we use a non dispersing phonon energy band
and set $w_q^{\alpha\gamma}$ to a momenta independent interaction coupling constant.

The increase in $\cQ$ as the system is driven out of equilibrium is due to the concomitant increase in the electronic temperature
and in the drift velocity. However in practice we find that since $\beta \ll 1$ the latter contribution is minute so that
the energy loss is dominated by heating.

\subsection{Numerical solution of the hydrodynamic equations}

Given the above expressions for the energy and momenta loss rates, we numerically solve the three coupled hydrodynamic equations.
In our calculations we use $\cD=20eV$ for the deformation potential, and $c=0.02v$ for the sound velocity.
\begin{figure}[h]
\includegraphics[width=0.9\linewidth]{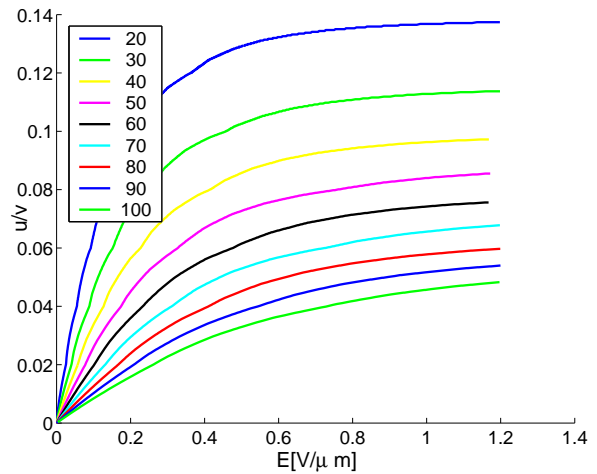}
\caption{Velocity saturation. The drift velocity normalized by the band velocity of graphene is plotted vs. electric field for
a series of lattice temperatures expressed in meV units.
}
\label{fig:current}
\end{figure}

\subsubsection{Clean limit}

In a clean system $\cP_i=0$ and the momentum loss is only due to the phonons.
In figure \ref{fig:current} the drift velocity is plotted \emph{vs.} electric field for $n=10^{13} cm^{-2}$ and $T_{\ty L}=25$ meV.
Clearly $du/dE$ is a monotonically decreasing function of the electric field.
At high fields $u$ approaches a saturation value. The highest fields in this figure correspond to an electronic temperature of $200$meV.
\begin{figure}[h]
\includegraphics[width=0.9\linewidth]{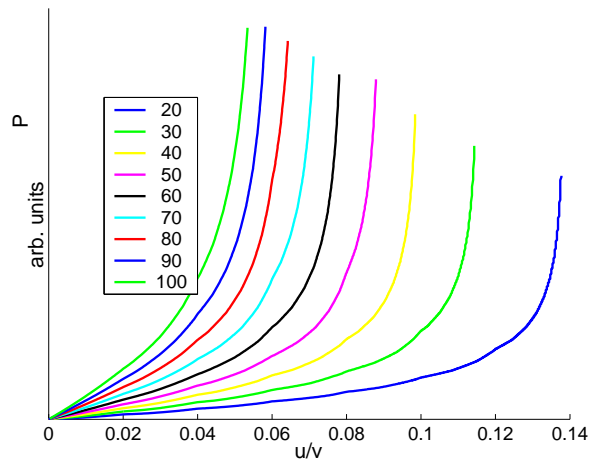}
\caption{Momentum loss rate as a function of drift velocity for a series of lattice temperatures expressed in meV units.}
\label{fig:Pdisperse}
\end{figure}

\begin{figure}
\includegraphics[width=0.9\linewidth]{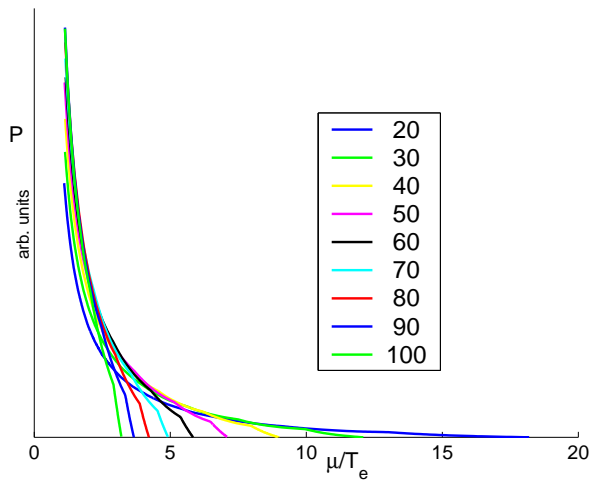}
\caption{The momentum loss rate as a function of $\mu/T_e$. Different curves correspond to
different values of the lattice temperature in meV units. Data collapses as the neutral regime is approached.
The drift velocity is limited by the ratio of the typical energy loss to the typical momentum loss averaged over
transitions, which is proportional to the Dirac band velocity.}
\label{fig:Pconverge}
\end{figure}

The hydrodynamic equations imply that the drift velocity is given by $\cQ/\cP$. To understand the saturation of the velocity we therefore study the dependence of $\cQ$ and $\cP$
on $u$. In figure \ref{fig:Pdisperse} we plot $\cP(u)$ for different values of the lattice temperature. The momentum loss increases with
$E$ with a sharp rise at the saturation velocity $u_{sat}(T_{\ty L})$
indicating that the saturation of the current is due to the enhanced scattering at high fields.
Similar behavior is obtained for the energy loss.
Interestingly, partial data collapse occurs when $\cQ$ and $\cP$ are plotted as a function of $\mu/T_e$ (see figure \ref{fig:Pconverge}),
demonstrating that a constant ratio of $\cQ$ and $\cP$ is reached as the neutral regime is approached.

We find that $u_{sat}$ is only weakly density dependent in the range $n=0.1-10\cdot 10^{12} cm^{-2}$. Therefore the saturation current $I_{sat}=e n u_{sat}$ is, to a good approximation,
linear in $n$.
We also find that for all values of the applied field  $P_o \ll P_a$.  Acoustic phonons therefore play an essential role in
the non linear {\em dc} electrical properties of graphene. The energy loss rate in doped graphene is dominated by acoustic phonons at low electronic
temperatures\cite{us}, however far from equilibrium it is the optical phonons that dominate $\cQ$.

In deriving $\cP_o$ and $\cQ_o$ we implicitly assumed that the optical phonons are thermalized.
Experimental and theoretical work has shown that the optical phonons may be far from equilibrium in hot carbon nanotubes,\cite{Song,Lazzeri}
and this is also a possibility in graphene.
The approximation we make here in setting the temperature of the optical phonons to the
temperature of the acoustic phonon bath $T_{\ty L}$ can be justified {\em a posteriori} by
our numerical results.  We find that $\cQ_o^{sp} \gg \cQ_o^{ind}$ far from equilibrium.
Any increase in the temperature of the optical phonon bath will
influence only $\cQ_o^{ind}$ and will thus have little effect on our
non-linear transport results.

\subsection{Influence of disorder   \label{sec_disorder}}

\begin{figure}[h]
\includegraphics[width=0.9\linewidth]{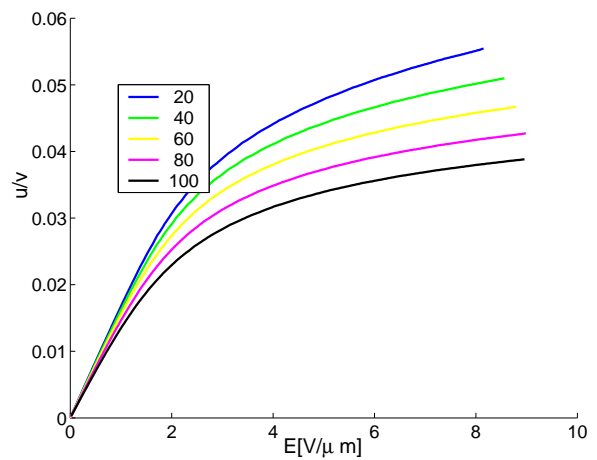}
\caption{Current as a function of electric field for $n=10^{13} cm^{-2}$ and
$n_i=10^{11} cm^{-2}$. The different curves correspond to different values of the lattice temperature in meV units.}
\label{fig:uDisorder}
\end{figure}

The presence of disorder leaves the form of the hydrodynamic equations unchanged, however it does increases the momentum loss
rate  $\cP$. Therefore,
for a given value of the electric field, we expect the drift velocity $u=\cQ/\cP$ to be reduced relative to its value in a clean system. In figure \ref{fig:uDisorder}
the drift velocity is plotted as a function of the electric field for a graphene sheet with an electronic density $n=10^{13} cm^{-2}$ and an
impurity concentration $n_i = 10^{11} cm^{-2}$. For the strongest fields $T_e = 280 meV$. Comparing figure \ref{fig:uDisorder} and figure \ref{fig:current}
clearly shows that the drift velocity is indeed reduced. Moreover we find no current saturation even at the highest fields.
Since disorder does not modify the energy loss, we expect saturation to occur only when $P_a \gg P_i$.

\section{Discussion       \label{sec_summary}}

At high electronic temperatures, rapid e-e collisions justify the hydrodynamic theory used here.
The high temperatures are characteristic of the non-linear response regime which is our principle focus,
but the hydrodynamic theory will also apply in the linear regime for sufficiently clean and hot samples.

In the linear response regime we find that inter-particle collisions have little effect on the resistivity when $\mu/T$ is large.
However at high temperatures, or alternatively at low densities, interactions between electrons increase the resistivity.
Recently Bolotin {\em et. al.} measured the resistivity in ultra-clean suspended graphene over a wide range of densities and temperatures\cite{Bolotin}.
Their experimental samples were close to the ballistic limit
in which the quantum contact resistances are a substantial fraction of the
overall resistance.  Surprisingly this study found that the phonon induced resistivity is density dependent at low densities.
Although our theory can not make quantitative predictions
in the ballistic regime since it is unable to account for the quantum resistance,
our results do suggest e-e interactions as a possible origin of this density dependence.

In the non-linear regime we find that $u=\cQ/\cP$: the drift velocity is given by the ratio between the energy loss rate and
the momentum loss rate. At strong fields
high electronic temperatures are responsible for a rapid increase of both $\cQ$ and $\cP$ that results in the saturation of $u$ at a velocity of the
order of $10^7 cm/sec$. The saturation velocity is only weakly density dependent therefore the saturation current $I_{sat} = e n u_{sat}$
increases, to a good approximation, linearly with density. The electronic temperature at which velocity saturation occurs increases with
the impurity concentration.  For even moderately disordered samples, electronic temperatures reach unphysically large values
before saturation occurs.

In this work we neglected the phonons of the substrate.  The interaction of Dirac quasi-particles with
substrate phonons can however be important, depending on the type of substrate and its distance from the graphene sheet \cite{Fratini,FuhrerSubstarteSiO2}.
Following  the prescription given in section \ref{sec_non_linear} the contribution of these phonons may straightforwardly be added to the the hydrodynamic
equations (\ref{DC_momentum},\ref{DC_energy}).
Strong coupling between the substrate phonon modes and the electrons may significantly lower the electronic temperature in which current saturation occurs.
If the coupling to substrate phonons is too strong ($\tau_{ph} < \tau_{ee}$), the hydrodynamic approach is invalidated.

Previous work attributed current saturation in carbon nanotubes\cite{Yao} to the
the sudden onset of momentum relaxation by zone-boundary optical phonons.
More recently Meric {\em et. al.} associated the saturation of the current in graphene based field-effect transistors
with surface phonons of the $SiO_2$ substrate\cite{meric}. In this work we find an additional mechanism for current saturation in which the increase of the drift velocity
is inhibited by high electronic temperature at strong fields.
The two mechanisms can be distinguished experimentally by the dependence of saturation velocity on carrier density which they predict.
The electron-heating mechanism leads to a saturation velocity which is $\sim 10\%$ of the Dirac velocity and weakly dependent on
carrier density.  The saturation current is therefore, to a good approximation, proportional to carrier density.
The phonon back-scattering mechanism, on the other hand, predicts a saturation velocity which is $\sim v \, \omega_{ph}/\mu$ and therefore
a critical current which varies as the square root of carrier density.

In real devices the top-gate lies in close proximity to the source-to-drain conduction channel.
The electric field in the channel is consequently strongly space dependent.
The hydrodynamic theory outlined above may be straightforwardly generalized to include such spatial variations. The out of equilibrium system
is then described by three hydrodynamic functions $\bm{u(r)}$, $T_e(\bm{r})$ and $\mu(\bm{r})$ whose values are determined by the
continuity equation, the momentum balance equation and the energy balance equation.

After completion of this paper we became aware of an interesting and closely related
recent study by Barreiro et. al..\cite{Barreiro}
This complementary work addresses high field transport in graphene both experimentally and theoretically.
The experimental results support our findings regarding the absence of current saturation in disordered graphene.
The theoretical analysis in Ref.(27) utilizes the Boltzmann equation, however it neglects e-e interactions
and the interactions of the Dirac quasi-particles with acoustic phonons.  As we explained above e-e
interactions are expected to be important when the electronic temperature is high, and interactions with acoustic
phonons are essential for the description of current saturation in clean graphene.

\acknowledgements

This work was supported by the Welch foundation and
by DARPA under contract FA8650-08-C-7838 through the CERA program.
The authors acknowledge helpful interactions with K.I. Bolotin and P. Kim.

\appendix

\section{Energy loss to acoustic phonons       \label{app_acoustical_phonons_Q}}

The energy loss rate due to the interactions of the electrons with acoustic phonons
\bea
\cQ_a &=& \frac{\pi g \cD^2}{2\rho c} \sum_{\bm{k,p}\alpha\gamma} \epsilon_{kp}^{\alpha\beta}(1+\alpha\gamma\cos\theta)
q f_{\bm{k}\alpha}^{\ty H} \nonumber \\
&\times& \left[ ( N_q+1 )\delta(\epsilon_{kp}^{\alpha\gamma}-\omega_q) + N_q \delta(\epsilon_{kp}^{\alpha\gamma}+\omega_q) \right] \label{Qa}
\eea
where $q=|\bm{k-p}|$, is evaluated to leading
order in $\beta=c/v$. Such an expansion is valid in the quasi-elastic scattering regime when $T_e>T_{\ty{BG}}$.

The energy conservation condition in (\ref{Qa}) inhibits inter-band transitions.
Furthermore it sets the value of the dummy variable $p$ to
\be
p_0 = k \pm 2\alpha k \beta|\sin(\theta/2)| + 2\beta^2 \sin^2(\theta/2)       \label{p_k}
\ee
where the positive sign relates to a phonon emission process and the minus sign to a phonon absorption
process. Hence to first order in $\beta$
\be
\delta(\epsilon_{kp}^{\alpha\gamma} \pm c|\bm{k-p}|) = \delta_{\alpha\gamma}\frac{\delta(p-p_0)}{v}\left[ 1 \mp \alpha\gamma|\sin(\theta/2)| \right]. \label{delta}
\ee
The total energy loss (\ref{Qa}) is a sum of $\cQ_a^{sp}$ the energy loss due to
spontaneous emission of phonons and $\cQ_a^{ind}$ the energy gain due to induced transitions.
For the latter Eqs.(\ref{Qa},\ref{p_k},\ref{delta}) imply
\be
Q^{ind} = \frac{8\cD^2 T_{\ty L}}{\rho v^2} \int \frac{k^3 dk d\theta_k d\theta_p}{(2\pi)^3} \sin^2\theta
\sum_\alpha \alpha f_{\bm{k}\alpha}.
\ee
Integrating first over $\theta_\bm{p}$ and then over $\theta_\bm{k}$ results in expression (\ref{Qa_ind})
for $\cQ_a^{ind}$. A similar derivation yields Eq.(\ref{Qa_sp}).

\end{document}